\documentclass[%
reprint,
superscriptaddress,
amsmath,amssymb,
aps,
prl,
]{revtex4-1}
\usepackage{graphicx}
\usepackage{dcolumn}
\usepackage{bm}
\usepackage{amsmath}
\usepackage{hyperref}
\hypersetup{colorlinks=true,linkcolor=blue,anchorcolor=blue,citecolor=blue,urlcolor=blue}
\usepackage{lineno}
\usepackage{enumerate}
\usepackage{color}
\usepackage{microtype}

\begin{document}
\preprint{APS/123-QED}

\title{From Filamentation to Stratification: Instability Dynamics in Scissors-Shaped Relativistic Beam-Plasma System}

\author{X. Liu}
\affiliation{ 
State Key Laboratory of Dark Matter Physics, Key Laboratory for Laser Plasmas, School of Physics and Astronomy, Shanghai Jiao Tong University, Shanghai 200240, China
}%
\affiliation{
Collaborative Innovation Center of IFSA (CICIFSA), Shanghai Jiao Tong University, Shanghai 200240, China
}

\author{D. Wu}
\email{dwu.phys@sjtu.edu.cn}
\affiliation{ 
State Key Laboratory of Dark Matter Physics, Key Laboratory for Laser Plasmas, School of Physics and Astronomy, Shanghai Jiao Tong University, Shanghai 200240, China
}%
\affiliation{
Collaborative Innovation Center of IFSA (CICIFSA), Shanghai Jiao Tong University, Shanghai 200240, China
}

\author{J. Zhang}
\email{jzhang@iphy.ac.cn}
\affiliation{ 
State Key Laboratory of Dark Matter Physics, Key Laboratory for Laser Plasmas, School of Physics and Astronomy, Shanghai Jiao Tong University, Shanghai 200240, China
}%
\affiliation{
Collaborative Innovation Center of IFSA (CICIFSA), Shanghai Jiao Tong University, Shanghai 200240, China
}
\affiliation{
\mbox{Institute of Physics, Chinese Academy of Sciences, Beijing 100190, China}}

\date{\today}

\begin{abstract}
Counter-streaming systems are a canonical model for beam-plasma instabilities, such as the filamentation instability, which is critical in high energy density physics. However, scenarios involving intersecting fast electron beams break the cylindrical symmetry inherent to such systems. Here, we introduce the scissors-shaped configuration, a fundamental multi-velocity-component system that captures this broken symmetry. Through theoretical analysis and particle-in-cell simulations, we reveal a dramatic shift in the instability dynamics: the system undergoes a stratification mode instead of filamentation. This mode is rapidly quenched by magnetic reconnection, leading to a quasi-stable state with magnetic energy two orders of magnitude lower than in the counter-streaming case. This discovery establishes a new principle of passive instability control via geometric configuration, offering a new perspective on beam-plasma interactions in astrophysics and inertial confinement fusion. The underlying physics is verifiable in upcoming multi-laser experiments.

\end{abstract}

\maketitle
$ Introduction $.---
The interaction of relativistic electron beams with dense plasmas is a fundamental process in high energy density physics (HEDP), underpinning phenomena in inertial confinement fusion (ICF) schemes like fast ignition (FI) \cite{r1,r2,r3,r4} and laboratory astrophysics \cite{r5,r6,r7,r8,r9}. In many of these scenarios, electromagnetic instabilities, such as filamentation instability, critically limit the transport efficiency and quality of fast electron beams \cite{r10,r11,r12,r13,r14}.

The study of beam-plasma instabilities has a rich history, advancing through theoretical, computational, and experimental efforts. Theoretically, since Weibel's seminal work \cite{r10}, analytical models have successfully described the linear phase of these instabilities. Key developments include the relativistic formulation of the Weibel instability by Yoon and Davidson \cite{r15}, explorations of the purely transverse Weibel instability relative to fast ignition by Silva \cite{r16}, and systematic investigations by Bret that clarified the role of oblique modes and compared the dominant parameter ranges of various instabilities \cite{r7,r17,r18}. These foundational theories have been extended to include collisional effects \cite{r19,r20} and have been complemented by a fully kinetic Vlasov model \cite{r21,r22,r23}. Computationally, particle-in-cell (PIC) simulations have provided crucial insights into the nonlinear dynamics, revealing filament merging as a primary mechanism driving fast electron beam deceleration and dissipation \cite{r24}, and elucidating the impact of various plasma parameters \cite{r12,r25,r26}. Theoretical and computational predictions have been corroborated by experimental observations, which have directly imaged beam filamentation and confirmed the presence of Weibel-generated magnetic fields \cite{r27,r28,r29}. However, the vast majority of these seminal studies have focused on the counter-streaming system with cylindrical symmetry, in which two symmetric plasmas counter-propagate, or a single fast electron beam propagates opposite to a background return current.

\begin{figure}[h]
  \includegraphics[scale=0.7]{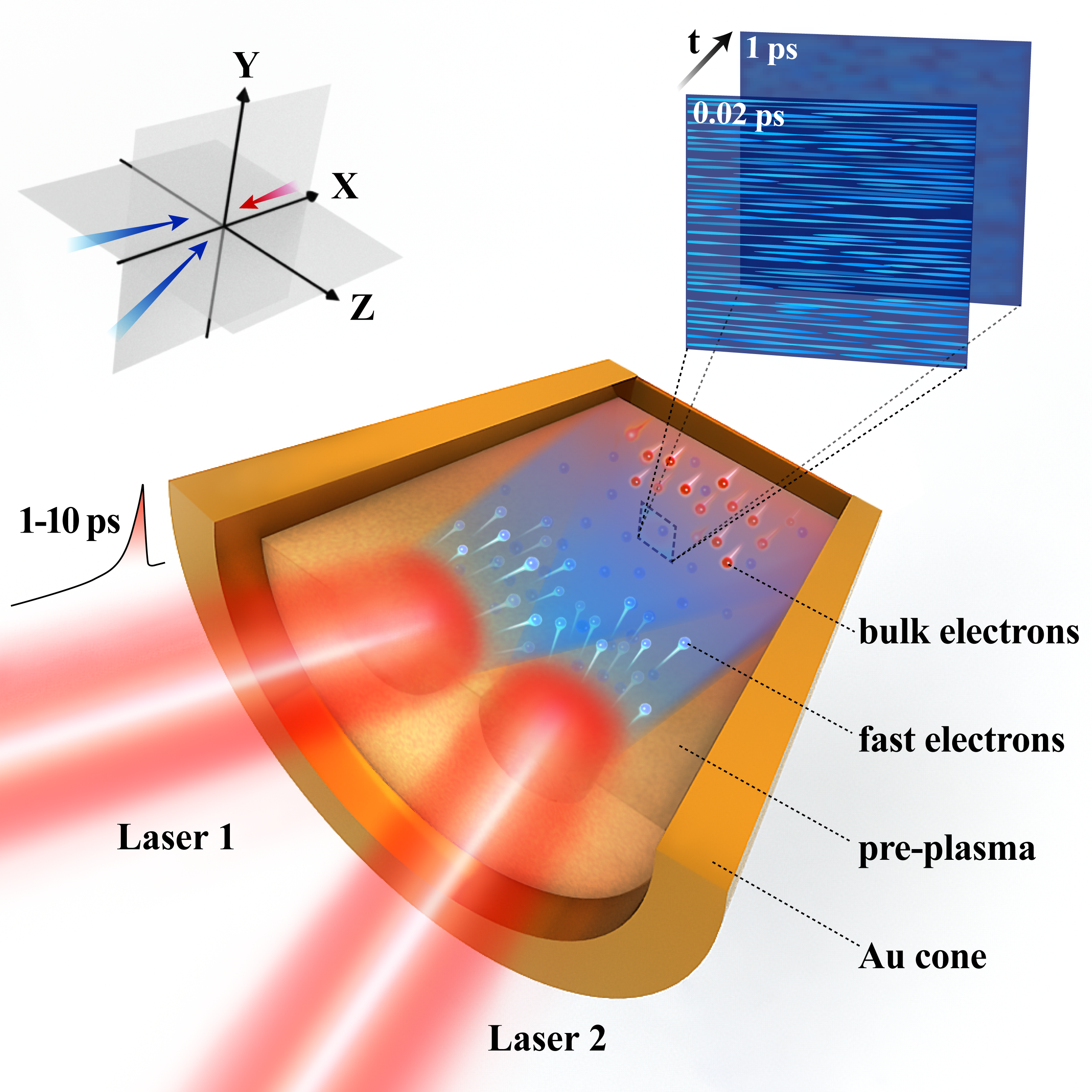}
  \caption{\label{fig:figure1} Schematic diagram of the scissors-shaped configuration.}
\end{figure}

A crucial and largely unexplored question is how the system evolves when cylindrical symmetry is broken. Such scenarios are not merely theoretical curiosities but are becoming increasingly relevant. For instance, in the overlapping relativistic electron jets in astrophysics, and in the double-cone ignition campaign \cite{r30}, which is upgrading the laser facility to employ two or more ultra-intense picosecond lasers to generate multiple fast electron beams for fast ignition as illustrated in Fig.\ \ref{fig:figure1}. In these scenarios, multiple fast electron beams intersect, forming a multiple-velocity-component system that cannot be transformed into a conventional counter-streaming system through reference frame transformation.

In this Letter, we address this question by introducing a canonical non-counter-streaming system: a scissors-shaped configuration formed by two intersecting fast electron beams and a background electron return current (with the electrical neutrality ensured by ions). Using theoretical analysis and PIC simulations, we demonstrate that breaking the cylindrical symmetry fundamentally alters the pattern of instability and the evolutionary behavior of the system. Instead of filamenting, the beams undergo a stratification mode, which is rapidly suppressed. We find this new configuration maintains excellent transverse beam uniformity for a duration of several picoseconds. In the nonlinear stage, we observe a secondary growth of magnetic field energy. This stage is termed the bulk cavitation stage, which ultimately terminates the transverse uniformity of the system. Our work reveals a novel mechanism for passively controlling beam-plasma instabilities and provides a new framework for understanding the interaction of multiple fast electron beams.

$Simulation$ $setups$.---Three sets of the most representative high-precision two-dimensional simulations were selected to investigate the evolution behavior of scissors-shaped configuration. First, simulation with a single fast electron beam injecting into the background (referred to as one-beam system) was conducted to offer a benchmark as the traditional counter-streaming system, in which the fast electron beam velocity is $\vec{v}_{b,1} = v_0 \hat{e}_x$. The second set is the scissors-shaped fast electron beams simulation (referred to as two-beam system), in which the velocities of the two symmetric fast electron beams are $\vec{v}_{b,2} = v_0 \cos 15^\circ \hat{e}_x \pm (v_0 \sin 15^\circ \cos 40^\circ \hat{e}_z + v_0 \sin 15^\circ \sin 40^\circ \hat{e}_y)$. The half-angle of 15$^\circ$ represents the initial direction of the fast electron beams relative to the x-axis. To represent the extreme case of further increasing the number of fast electron beams, an infinite-beams simulation (referred to as infinite-beam system) was selected as the third set. In this simulation, the fast electron beam velocity is isotropic in the z-y plane, given by $\vec{v}_{b,\inf} = v_0 \cos 15^\circ \hat{e}_x + v_0 \sin 15^\circ \hat{e}_\perp$. The x-direction is referred to as the longitudinal direction, while the z-y plane is designated as the transverse plane. Our two-dimensional simulations are performed in the z-y plane, capturing the evolution at a cross-section of the propagating fast electron beams.

\begin{figure*}[ht]
  \includegraphics[scale=0.542]{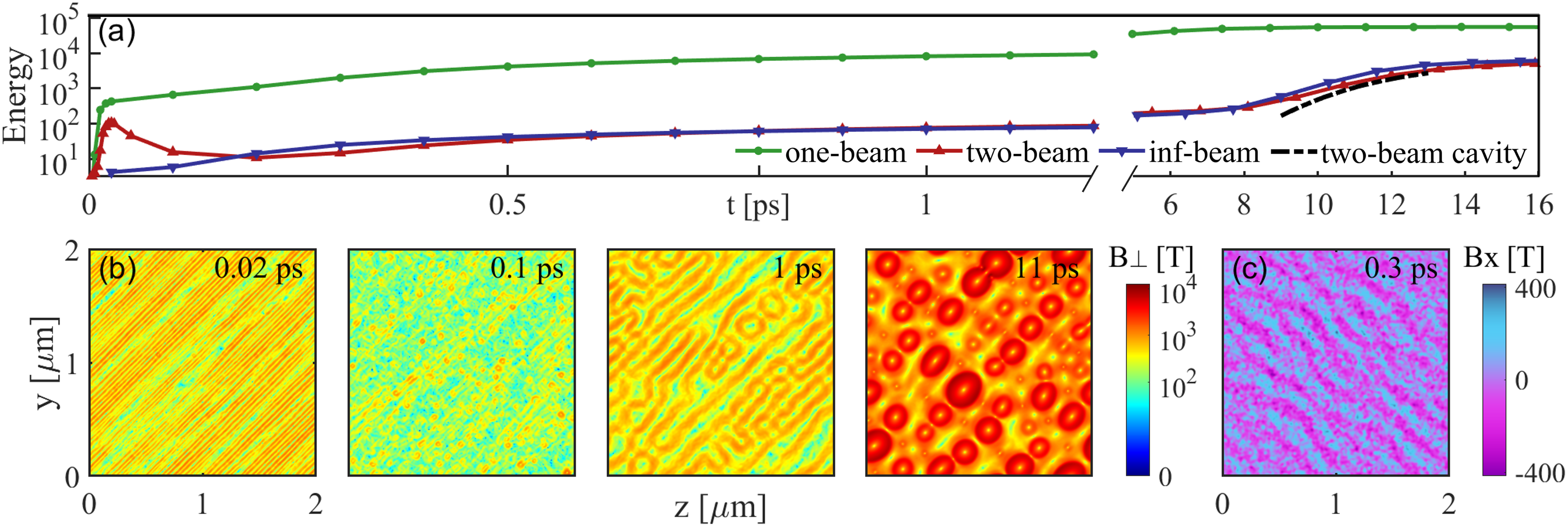}
  \caption{\label{fig:figure2} (a) Temporal evolution of magnetic field energy in three cases, with black dashed line indicating the magnetic field energy inside the cavities during the bulk cavitation stage of the two-beam system. The time axis includes a break to emphasize the initial and late-time evolution; (b) The magnetic field in the z-y plane $B_\perp$ of the two-beam system at $t = 0.02$, $0.1$, $1$, and $11$ ps; (c) The magnetic field in the x-direction $B_x$ of the two-beam system at $t = 0.3$ ps.}
\end{figure*}

In each set of simulation, the fast electron energy is 10 MeV ($\gamma = 20$) to make the relativistic effects more pronounced, and the total density of the fast electron beams is $n_b = 2n_c=2.4\times10^{21}$ cm$^{-3}$. The background electron density is $n_p = 500n_c$, which is a typical value around Au cone tip in fast ignition. The background electron velocity is $\vec{v}_p = -(n_b/n_p) v_0\hat{e}_x$ for the one-beam system and $\vec{v}_p = -(n_b/n_p) v_0 \cos 15^\circ\hat{e}_x$ for the two-beam and infinite-beam systems to ensure current neutralization. Au ions, which lack directed velocity, have an ionization degree $Z = 10$, a mass $m_i = 361682 m_e$, and a number density $n_i = 50.2 n_c$. Both ions and electrons have an initial temperature of 500 eV, and each particle is uniformly distributed within the simulation box. The simulation employs a uniform grid spanning the $z$-$y$ plane, with $\mathrm{N}_z \times \mathrm{N}_y = 200 \times 200$ cells. 
Each grid cell has edge lengths $\Delta z = \Delta y = 0.01~\mu\mathrm{m}$, and contains 400 background electrons, 200 fast electrons, and 100 background ions. The high number of particles per cell (ppc) ensures that the transverse velocity of fast electrons is nearly isotropic in each cell of the infinite-beam system. While fully three-dimensional simulations are computationally extremely challenging, our two-dimensional simulations with periodic boundary conditions focus on a small transverse cross-section of the system which captures the essential physics. In the supplementary material, we present a convergence analysis using larger simulation sizes, demonstrating that doubling the side length in the z-y plane yields consistent results with those reported here. Both collisionless and collisional scenarios were investigated, and our main focus is on the collisionless scenario, with the collisional scenario detailed in the supplementary material \cite{r31}.

$Linear$ $stage$.---Fig.\ \ref{fig:figure2}(a) illustrates the temporal evolution of the total magnetic field energy for the three sets of simulations. The evolution of magnetic field energy in the two-beam system initially approximates that of the one-beam system, but after a short period, it declines sharply and becomes essentially consistent with the infinite-beam system. Both two-beam and infinite-beam systems exhibit significantly lower magnetic field energy compared to the one-beam system, indicating a substantial suppression of electromagnetic instability. Then around $t = 8$ ps, an unusual secondary growth occurs in these two systems.

\begin{figure}
  \includegraphics[scale=0.5]{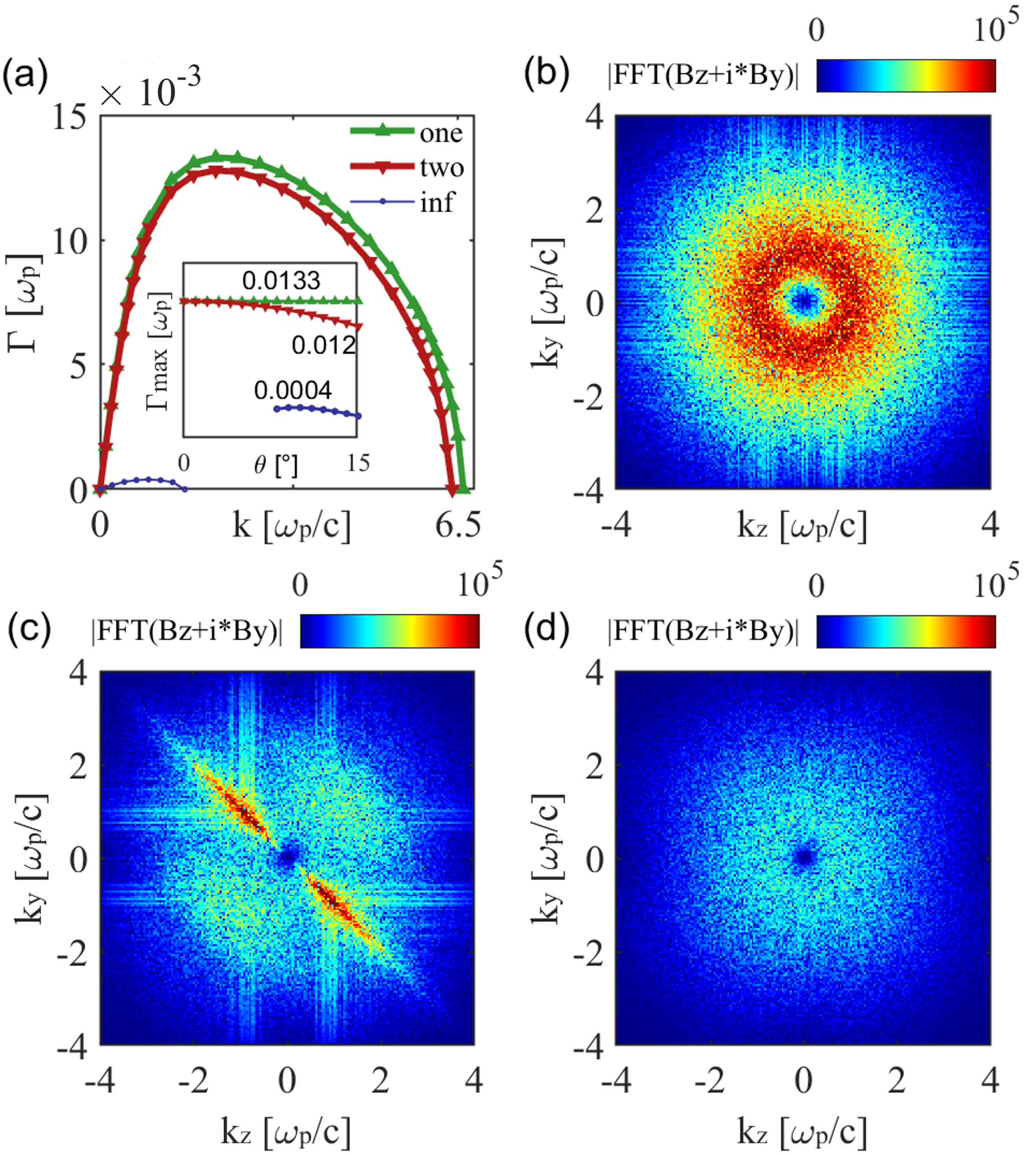}
  \caption{\label{fig:figure3} (a) Dispersion relation curves for the three systems. The inset shows the variation of the maximum growth rate with the initial incident angle of the fast electron beams for the three systems. (b)(c)(d) Two-dimensional Fourier analysis results of $(B_z + i \cdot B_y)$ for one-beam, two-beam, and infinite-beams systems at the linear stage ($t = 0.02$ ps), respectively.}
\end{figure}

During the linear phase, the dispersion relations for the three systems can be derived using linear theory \cite{r6,r17,r18,r31}. The maximum instability growth rates for the one-beam, two-beam, and infinite-beam systems, denoted as $\Gamma_1$, $\Gamma_2$, and $\Gamma_{\text{inf}}$ respectively, exhibit a trend of $\Gamma_1>\Gamma_2\gg\Gamma_{\text{inf}}$. This trend matches the simulation results well, as shown in Fig.\ \ref{fig:figure2}(a) and Fig.\ \ref{fig:figure3}(a). The two-dimensional Fourier analysis results of $(B_z + i \cdot B_y)$ for the three systems at $t = 0.02$ ps are presented in Fig.\ \ref{fig:figure3}(b)-(d). In contrast to the other two systems, the broken symmetry in the two-beam system dictates a selective excitation of modes, whose wave vectors are concentrated along the new symmetry axis of the system (perpendicular to the plane of the fast electron beams' initial velocities), corresponding to a stratification mode. The excited magnetic field structure is not the traditional toroidal shape but a series of parallel straight lines with alternating directions, as shown in Fig.\ \ref{fig:figure2}(b), causing the current density of the fast electron beams and background electrons to form a similar layered structure. The small window in Fig.\ \ref{fig:figure3}(a) shows that as the incident angle decreases, the theoretical growth rate in the two-beam system converges to that in the one-beam system, providing a crucial self-consistency check for the theory. However, for infinite-beam system, due to the different topologies of the initial distribution function, the mathematical derivation results cannot converge with small incident angles, which is discussed in the supplementary material.

\begin{figure*}
  \includegraphics[scale=0.93]{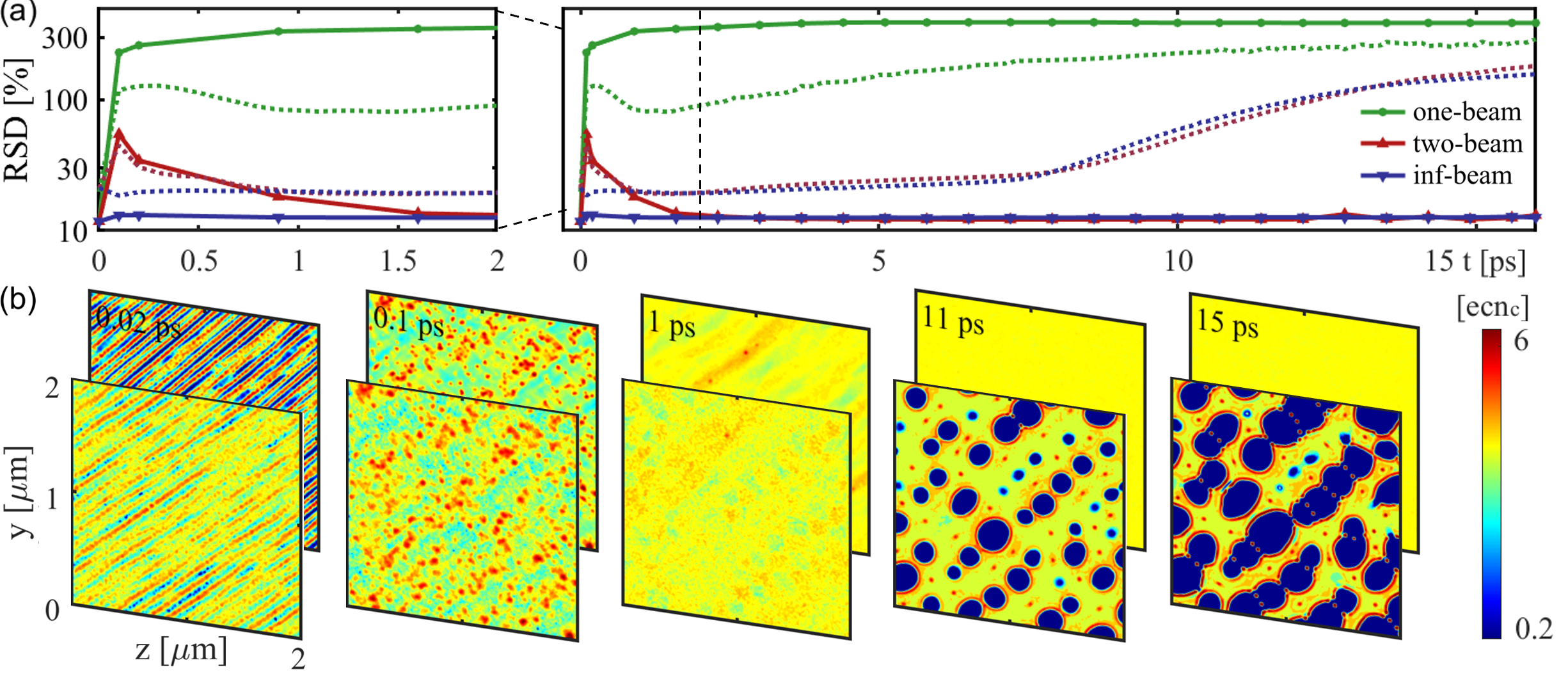}
  \caption{\label{fig:figure4} (a) Transverse inhomogeneity (RSD) of the fast electron (solid lines) and background electron (dashed lines) current densities for the three systems; (b) The transverse distribution of the background current density $j_{p,x}$ (front) and the beam current density $-j_{b,x}$ (behind). Each column corresponds to $ t = 0.02$, $0.1$, $1$, $11$ and $15$ ps, respectively.}
\end{figure*}

Although the growth rate $\Gamma_2$ of the stratification mode is close to $\Gamma_{1}$ of the filamentation instability, within a mere 0.03 ps, the stratification mode is suppressed due to collisionless magnetic reconnection, leading to a sharp decline in magnetic field energy. The layered structures of the magnetic field and fast electron beams quickly disintegrate, resulting in a disordered state at $t =$ 0.1 ps \cite{r31}.

Additionally, at $t\sim0.3$ ps, due to the anisotropy of the fast electron beams' velocity in the z-y plane, Weibel-like instability and the magnetic field in the x-direction are excited, as shown in Fig.\ \ref{fig:figure2}(c). Although this instability exists for a very short time, the residual magnetic field in the x-direction continues to drive the two-beam system toward the infinite-beam system, which reaches transverse velocity isotropy around $t = 8$ ps \cite{r31}.

The magnetic field structure in the two-beam system is similar to the infinite-beam system for several picoseconds after the suppression of the stratification mode, forming an interconnected maze-like structure. At $t\sim6$ ps, the magnetic field energies of all three systems temporarily reach a quasi-steady state, with the magnetic field energy of the two-beam and infinite-beam systems being about two orders of magnitude lower than that of the one-beam system, demonstrating a significant suppression effect on electromagnetic instabilities.

$Nonlinear$ $stage$.---For both the two-beam and infinite-beam systems, around $t = 8$ ps, the magnetic field energy underwent a secondary growth. In this stage, the systems' behavior could no longer be described by the linear theory of electromagnetic instability. Due to the isotropy of the fast electrons transverse velocity in two-beam system around $t = 8$ ps, the two-beam and infinite-beam systems behave similarly and are hereafter discussed collectively.

Around $t = 8$ ps, the connected magnetic field structure is disrupted by collisionless magnetic reconnection, forming closed magnetic field vortices. The Lorentz force within counterclockwise magnetic vortices expels background plasma (electrons and ions), carving out a series of low-density cavities and leading us to designate this phase as the bulk cavitation stage.

Simulations revealed that during the bulk cavitation stage, the density of background electrons and ions within the cavities decreased to the level of the beam electron density ($\sim1 n_c$), causing the background electron current density in the cavities to nearly vanish. However, due to the high transverse momentum, the net Lorentz force integrated along a fast electron's trajectory through a cavity is nearly zero, leaving their transverse momentum largely unperturbed. Thus the fast electron beams maintained a nearly uniform distribution in the z-y plane, as illustrated in Fig.\ \ref{fig:figure4}(b). The complete evolution of the fast electron beam and the background electron current density is visualized in Movie S1 in the Supplemental Material \cite{r31}.

It is found that the magnetic field strength inside a cavity is roughly proportional to the distance from its center, while the equivalent cavity radius grows approximately linearly with time \cite{r32}. The evolution of the total magnetic field energy within the cavities $E_{in}$ for the two-beam system is plotted in Fig.\ \ref{fig:figure2}(a) (black dashed line). To explain this behavior, we developed a simplified hydrodynamic model, which predicts that the magnetic energy should grow as the fourth power of time, i.e., $E_{\text{in}} \sim A(t-t_0)^4$. To verify this prediction, we performed a power-law fit to the simulation data during the bulk cavitation stage. The resulting exponent consistently falls within the range of 3.95 to 4.05, in excellent agreement with our theoretical model. This correspondence suggests that the secondary growth of magnetic energy is well-captured by our cavitation model, rather than being driven by other potential instabilities.

$Transverse$ $uniformity$.---The degree of transverse spatial inhomogeneity of the fast electron beams and background electrons can be represented by the relative standard deviation (RSD) of the current density:

\begin{equation}
\begin{gathered}
\rm{RSD}=\sqrt{\frac{1}{S}\int\int\frac{(j_x(z,y)-<j_x>)^2}{<j_x>^2}dzdy}
\end{gathered}
\end{equation}

Fig.\ \ref{fig:figure4}(a) illustrates that as the stratification mode of the two-beam system is suppressed, the fast electron beam current density returns to a state closely resembling the initial uniform state. The RSD of the fast electron beams in both the two-beam and infinite-beam systems remains around $10\%$, and the uniformity of the background electrons is significantly improved compared to the one-beam system before $t = 8$ ps. In contrast, due to filamentation instability, the RSD of the fast electron beam in the one-beam system rapidly exceeds $300\%$, and the RSD of the background electrons surpasses $100\%$.

During the bulk cavitation stage, the background RSD in the two-beam and infinite-beam systems rapidly increases, resembling a filamentation-like instability driven solely by the background return current. This stage signifies the end of the system's transverse uniformity, indicating that our method for enhancing the transverse uniformity of the fast electron beam is effective for approximately 8 ps, which is sufficient for typical picosecond laser pulses in fast ignition. While the fast electron beams remain transversely uniform in our two-dimensional simulations, a crucial consideration in a realistic three-dimensional scenario is their longitudinal transport. As the initial electron population propagates out of the system along the x-direction, it is replaced by a continuous influx of new electrons. These newly introduced beams will encounter the pre-formed background cavities and strong magnetic fields, making them highly susceptible to filamentation, particularly for electrons with lower energy.

\begin{figure}
  \includegraphics[scale=0.4]{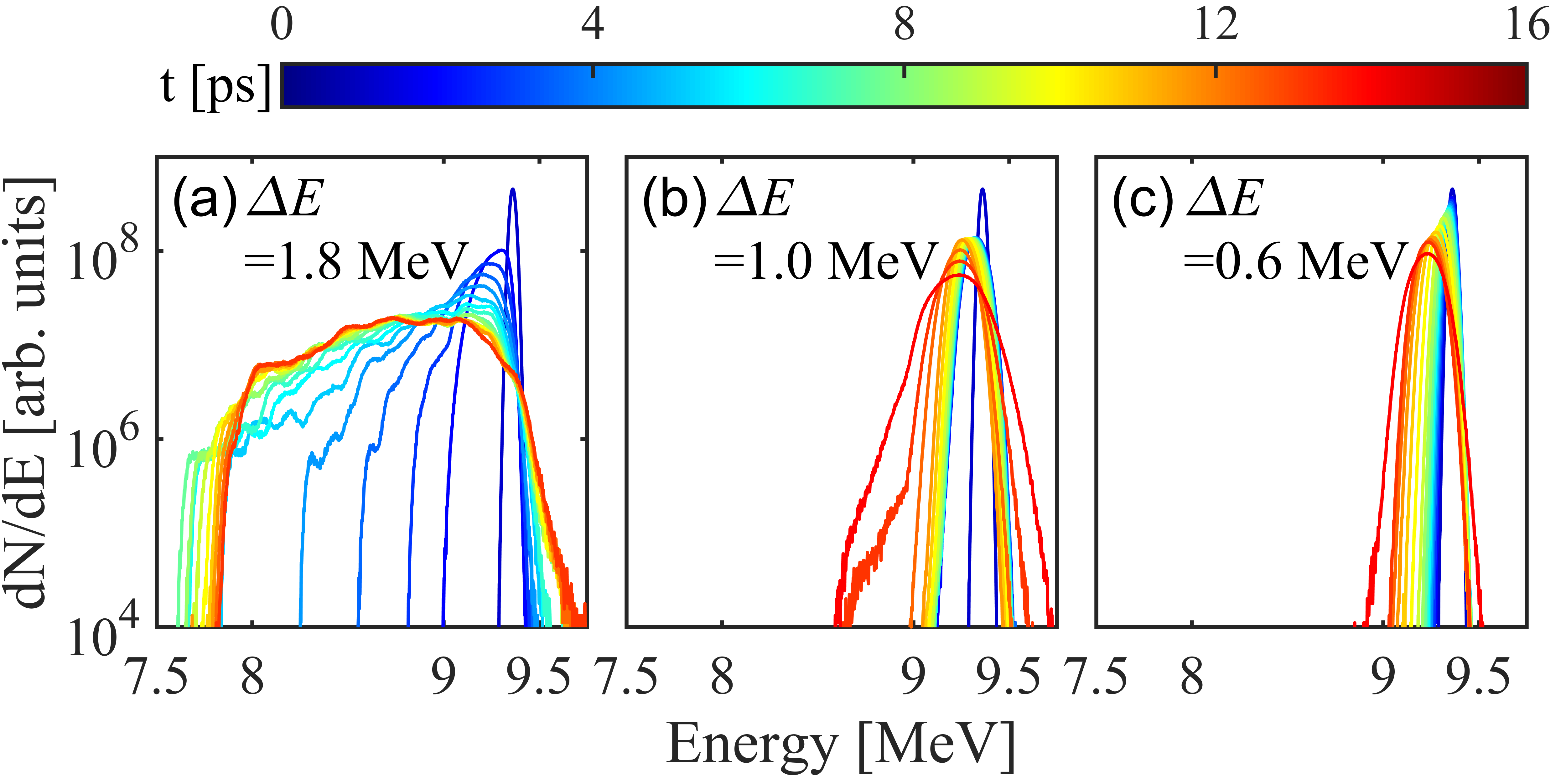}
  \caption{\label{fig:figure5} Fast electron energy spectrum of (a) one-beam, (b) two-beam and (c) infinite-beam system. The displayed energy values are the kinetic energy of the fast electrons, calculated by subtracting their rest energy. $\Delta E$ is the final full width at half maximum.
}
\end{figure}

$Energy$ $spectrum$.---By comparing the evolution of fast electron beam energy spectrum in the three systems depicted in Fig.\ \ref{fig:figure5}, we observed that both the two-beam and infinite-beam systems effectively mitigate the thermalization and energy loss of fast electron beams caused by filamentation instability. Filamentation instability leads to a gradual broadening and leftward shift of the fast electron beam energy spectrum in the one-beam system, with the broadening reaching approximately 2 MeV. Conversely, in the two-beam and infinite-beam systems, the fast electron energy remains relatively stable, with only a minimal broadening of the energy spectrum over the picosecond timescales relevant to fast ignition. This indicates that the instabilities in multi-velocity-component systems have less impact on the deceleration and heating of fast electrons.

$Discussion$ $and$ $conclusion$.---
In conclusion, we have demonstrated that breaking the cylindrical symmetry in a relativistic beam-plasma system fundamentally alters its instability dynamics. By employing a novel scissors-shaped configuration, the dominant instability shifts from filamentation to a stratification mode, which is rapidly quenched by magnetic reconnection. This yields a state of strong magnetic field suppression and remarkably high beam uniformity persisting for several picoseconds. The mechanism remains effective even in collisional cases \cite{r32}. This study establishes a new principle: geometric configuration can serve as a powerful passive tool for controlling beam-plasma instabilities, enriching our understanding of multi-velocity-component plasma systems. While our two-dimensional simulations provide a proof-of-principle, the evolutionary behavior of multi-velocity-component systems in three dimensions remains a critical open question for future investigation. Nevertheless, our work offers a new perspective for interpreting phenomena in overlapping astrophysical jets and for designing future multi-laser HEDP experiments, with key predictions on the initial instability evolution being testable at upcoming dual-beam laser facilities.

This work was supported by the Strategic Priority Research Program of Chinese Academy of Sciences (Grants No. XDA25050500 and No. XDA25010100), the National Natural Science Foundation of China (Grants No. 12075204), and the Shanghai Municipal Science and Technology Key Project (No. 22JC1401500). D. W. thanks the sponsorship from Yangyang Development Fund.

\nocite{*}

\bibliography{Reference}

\end{document}